\documentclass[aps,pre,showpacs]{revtex4}
\usepackage{graphicx}
\usepackage{epsfig}
\begin{document}
\title{Emergence of patterns in driven and in autonomous
spatiotemporal systems}
\author{ M. G. Cosenza, M. Pineda and A. Parravano}
\affiliation{Centro de Astrof\'{\i}sica Te\'orica,
Facultad de Ciencias Universidad de Los
Andes, M\'erida, Apartado Postal 26, M\'erida~5251, Venezuela.}
\date{  }
\begin{abstract}
The relationship between a driven extended system and an autonomous
spatiotemporal system is investigated
in the context of coupled map lattice models.
Specifically, a locally coupled map lattice subjected to an
external drive is compared to a coupled map system with similar local
couplings plus a global interaction.
It is shown that, under some conditions, the emergent patterns in both
systems are analogous. Based on the knowledge of
the dynamical responses of the driven lattice, we present a
method that allows the prediction of parameter values for the emergence of
ordered spatiotemporal patterns in a class of coupled
map systems having local coupling and general forms of global interactions.
\end{abstract}
\maketitle

The phenomenon of pattern formation induced by external forcing on
spatiotemporal systems, such as chemical reactions
\cite{Swinney,Hudson1,Kapral} or granular media \cite{Swinney1,
Swinney2,Metcalf}, have received much attention. Similarly, there
has been recent interest in experimental investigations of
spontaneous pattern formation and emergence of collective
behaviors in spatially extended systems of interacting dynamical
elements, such as one-dimensional arrays of electrochemical
oscillators \cite{Ring}, chemical and hydrodynamical systems with
global coupling \cite{Hudson2,Yamada}, and populations of chaotic
electrochemical cells having both local and global interactions
\cite{Hudson3}. In this context, coupled map lattices have
provided fruitful theoretical models for studying and predicting a
variety of dynamical processes, including pattern formation, in
spatiotemporal chaotic systems possessing different kinds of
interaction topologies, such as local couplings, regular
geometries, inhomogeneous and disordered networks, and global
coupling \cite{Chaos}.

In this article we investigate the relationship between forced
spatiotemporal systems and autonomous dynamical systems possessing
both local and global interactions in the framework of coupled map
lattices. We analyze the emergence of ordered patterns in forced
spatiotemporal systems by using a model of a coupled map lattice
subjected to an external drive. We show that, under some
circumstances, this system is analogous to an autonomous coupled
map system with both local and global interactions. Our approach
is motivated by the observation that a globally coupled map system
behaves similarly to a single map subjected to an external drive
and that this analogy may be used to describe the formation of
dynamical clusters in globally coupled maps \cite{PC}.

As a model of an autonomous spatiotemporal system, we consider the
following coupled map network possessing both, local and
global interactions \cite{Kaneko},
\begin{equation} 
\label{stcm}
  x_{t+1}^i= 
 (1-\epsilon_2)f(x_t^i)  
+\frac{\epsilon_1}{2} \left[(f(x_t^{i+1})+f(x_t^{i-1})-2f(x_t^i) \right] 
      +\epsilon_2 H \left( x^1_t,x^2_t,x^3_t,\ldots,x^N_t \right),
\end{equation}
which can be compared to a one-dimensional coupled map lattice subjected to
a uniform external drive,
\begin{equation}
\label{drnw}
  s^i_{t+1}=(1-\epsilon_2)f(s_t^i)
  + \frac{\epsilon_1}{2}\left[(f(s_t^{i+1})+f(s_t^{i-1})-2f(s_t^i)\right]
  +\epsilon_2 L_t \,.
\end{equation}
In Eq. (\ref{stcm}), $x_t(i)$ gives the state of element $i$
$(i=1,2,\ldots,N)$ at discrete time $t$; $N$ is the size of the
system;  $f(x)$ describes the (nonlinear) local dynamics;
$\epsilon_1$  and $\epsilon_2$ are the local and global coupling
parameters, respectively; and
$H(x^1_t,x^2_t,x^3_t,\ldots,x^N_t)$ is a global
coupling function of $N$ variables assumed to be invariant to
argument permutations; that is
$H(\ldots,x_t^i,\ldots,x_t^j,\ldots)=H(\ldots,x_t^j,\ldots,x_t^i,\ldots)$;
$\forall \, i,j$. This property of the global coupling function
ensures that, at any time, each element of the lattice is
subjected to the same influence of the global coupling term.
Thus the interaction $H$ provides a global feedback.
As a specific example, we shall take the usual mean field global
coupling $H= \frac{1}{N}\sum_{i=1}^N f(x_t^i)$; although the
analysis presented here is applicable whenever a permutable
coupling function of $N$ variables appears in the autonomous
spatiotemporal dynamical system Eq. (\ref{stcm}).
In Eq.~(\ref{drnw}), $s_t^i$ is the state of element $i$ in the driven
lattice, $f(s_t^i)$ is the same local dynamics as in Eq.
(\ref{stcm}), $\epsilon_1$ measures the local coupling,
$\epsilon_2$ represents the coupling strength to the external
forcing, and $L_t$ is the uniform driving term. In general, $L_t$
may be any function of time. We assume periodic boundary
conditions and a quadratic map $f(x)=1-r x^2$ as local dynamics in
both Eq. (\ref{stcm}) and Eq. (\ref{drnw}).

The analogy between an autonomous coupled map system having both
local and global interactions and a uniformly driven lattice
arises because in the former system (Eq.\ref{stcm}) all the
elements in the network are affected by the global coupling in
exactly the same way at any time, and therefore the behavior of
any map $x_t^i$ in Eq.(\ref{stcm}) is equivalent to the behavior
of any element in the driven lattice (Eq. \ref{drnw}) with $L_t=H$
and initial conditions $s_o^i=x_o^i$. Additionally, if the
autonomous coupled map system in Eq.(\ref{stcm}) reaches an
ordered spatiotemporal pattern, its corresponding global coupling
function $H$ follows an ordered motion. Thus the associated driven
lattice, Eq.~(\ref{drnw}), with a periodic forcing $L_t$ should
exhibit an ordered spatial pattern similar to that of the coupled
map system having both local and global interactions. In
particular, periodic drives resulting in periodic patterns in the
driven lattice Eq.(\ref{drnw}) may be employed to predict the
emergence of periodic patterns in autonomous systems described by
Eq.(\ref{stcm}), regardless of the specific functional form of the
permutable global coupling $H$ and without doing direct
simulations on the autonomous system.

Let us analyze the dynamical response of the driven coupled map lattice
subjected to periodic forcing. For a
drive having period $T$, we denote the sequence of $T$
values adopted by $L_t$ as $\{L_1,L_2,\ldots,L_T\}$. For some parameter values
and initial conditions, the resulting
dynamics of the driven system (Eq.\ref{drnw}) may reach an ordered
spatiotemporal pattern corresponding to lattice configurations
that are periodic in both, space and time. The spatial wavelength
$k$ and the temporal period $p$ of the dynamics of the elements
are defined by the relations $s_t^i=s_t^{i+k}$ and
$s_t^i=s_{t+p}^i$, respectively. In these cases it suffices to
consider only the dynamics of the $k$ coupled elements forming
a wavelength, and therefore the behavior of driven lattice
(Eq.\ref{drnw}) can be analyzed by considering a reduced driven
system of $k$ coupled maps.

An ordered spatiotemporal pattern having wavelength $k$ and period
$p$ emerging in a periodically driven lattice can be characterized
by the $k \times p$ matrix
\begin{equation}
S(k,p)= \left(
\begin{array}{ccc}
\sigma_1^1 & \ldots & \sigma_1^k \\
\vdots & \ddots & \vdots \\
\sigma_p^1 & \ldots & \sigma_p^k \\
\end{array}
\right) \, ,
\end{equation}
where the $m\mbox{th}$ column contains the consecutive $p$ values
of the periodic response of an element $m$ belonging to a spatial
wavelength, $(m=1,2,\ldots,k)$; and the $n\mbox{th}$ row displays
the values of all the $k$ elements in a wavelength (a snapshot) at
the cyclic time step $n$, $(n=1,2,\ldots,p)$. The driven lattice
may reach different asymptotic ordered spatiotemporal patterns
$S(k,p)$ depending on the initial conditions; i.e. multistability
is possible.

Once a pattern  $S(k,p)$ appears in the driven lattice for some
values of the parameters, it must satisfy the following set of $k
\times p$ nonlinear algebraic equations,
\begin{equation}
\label{kl}
  \sigma^m_{n+1}=(1-\epsilon_1-\epsilon_{2})f(\sigma_n^m)+
\frac{\epsilon_1}{2}[(f(\sigma_n^{m+1})+f(\sigma_n^{m-1})]
  +\epsilon_2 L_n.
\end{equation}
In general, for emergent periodic patterns the ratio $p/T=\nu$ is
a natural number that characterizes the resonance between the period
of the driving term and the resulting  period of the lattice.
Thus, the sequence of values $\{L_1,L_2,\ldots,L_T\}$ repeat
themselves $\nu$ times in the right hand side of Eqs.(\ref{kl}).
In addition, the presence of  symmetries in a spatiotemporal pattern
may reduce the number of independent variables in the above set of equations.
Eqs.~(\ref{kl}) may yield several sets of solutions for the orbits
$\sigma_n^m$, however only the stable ones will be observed as
asymptotic patterns in the driven lattice.
As examples, consider the following
patterns in the driven lattice, characterized by the given matrix and
satisfying the indicated relations, corresponding to Eqs.(\ref{kl}):\\
(1) Frozen ($p=1$) wavelength $k=2$,  with constant drive $L_t=\{L_1\}$,
\begin{equation}\label{K2P1}
S(2,1)=  \left( a \; b \right) \, ;
\end{equation}
\begin{equation}\label{21}
\begin{array}{ll}
   a= &  (1-\epsilon_1-\epsilon_2)f(a)+\epsilon_1 f(b)+\epsilon_2 L_1 \\
  b=&  (1-\epsilon_1-\epsilon_2)f(b)+\epsilon_1 f(a)+\epsilon_2 L_1 \, .
\end{array}
\end{equation}
(2) Out of phase, wavelength $k=2$ and period $p=2$, with constant drive
$\{L_1\}$,
\begin{equation}\label{K2P2}
S(2,2)=
 \left(
\begin{array}{cc}
a  & b \\
b & a
\end{array}
\right) \, ;
\end{equation}
\begin{equation}\label{22}
\begin{array}{ll}
   a= &  (1-\epsilon_1-\epsilon_2)f(b)+\epsilon_1 f(a)+\epsilon_2 L_1 \\
  b=&  (1-\epsilon_1-\epsilon_2)f(a)+\epsilon_1 f(b)+\epsilon_2 L_1 \, .
\end{array}
\end{equation}
(3) Wavelength $k=2$ and period $p=4$, with period-two drive $\{L_1,L_2\}$,
\begin{equation}\label{K2P4}
S(2,4)=  \left(
\begin{array}{cc}
a  & b \\
c & d  \\
b & a \\
d & c
\end{array}
\right) \, ;
\end{equation}
\begin{equation}\label{24}
\begin{array}{ll}
  c= &  (1-\epsilon_1-\epsilon_2)f(a)+\epsilon_1 f(b)+\epsilon_2 L_1  \\
  d= &  (1-\epsilon_1-\epsilon_2)f(b)+\epsilon_1 f(a)+\epsilon_2 L_1 \\
  a= &  (1-\epsilon_1-\epsilon_2)f(d)+\epsilon_1 f(c)+\epsilon_2 L_2 \\
  b= &  (1-\epsilon_1-\epsilon_2)f(c)+\epsilon_1 f(d)+\epsilon_2 L_2 \, .
\end{array}
\end{equation}
(4) Wavelength $k=3$ and period $p=3$, with constant drive $\{L_1\}$,
\begin{equation}\label{K3P3}
S(3,3)=  \left(
\begin{array}{ccc}
a  & b & c\\
b  & c & a \\
c  & a & b
\end{array}
\right) \, ;
\end{equation}
\begin{equation}\label{33}
\begin{array}{ll}
  a=&  (1-\epsilon_1-\epsilon_2)f(c)+\frac{\epsilon_1}{2}\left[f(b)+ f(a)\right]+
       \epsilon_2 L_1  \\
  b=&  (1-\epsilon_1-\epsilon_2)f(a)+\frac{\epsilon_1}{2}\left[f(c)+ f(b)\right]+
       \epsilon_2 L_1  \\
  c=&  (1-\epsilon_1-\epsilon_2)f(b)+\frac{\epsilon_1}{2}\left[f(a)+ f(c)\right]+
       \epsilon_2 L_1 \, . \\
\end{array}
\end{equation}

For the above examples, the orbits  $a,b,c,d$ can be obtained as
functions of the parameters $\epsilon_1, \epsilon_2, r$, and the
values $L_1$ and $L_2$.

In practice, if we are searching for the orbits $\sigma_n^m$ in a
specific stable pattern $S(k,p)$, we can construct a bifurcation
diagram of a periodically driven lattice of $N$ coupled maps ($N$
multiple of $k$) as a function of some parameter of the system,
and look for windows where that pattern appears in the diagram.
This may require exploring for appropriate initial conditions.
Figures~1(a)-(d) show bifurcation diagrams of driven lattices as a
function of either $\epsilon_1$ or $\epsilon_2$, where the
patterns given in examples (1)-(4), additionally to other
spatiotemporal states, arise. Fig.~1(a) shows the bifurcation
diagram of a lattice subjected to a constant drive $L_1=0.39$ as a
function of the coupling parameter of the drive $\epsilon_2$, with
fixed local coupling $\epsilon_1=0.54$. As $\epsilon_2$ is varied,
the steadily driven system exhibits several spatiotemporal states,
such as synchronization (both chaotic and periodic), a pattern
$S(2,1)$,  and quasiperiodic $k=2$ wavelength. The stationary
orbits $a$ and $b$ (for either the odd or even elements of the
lattice) in the pattern $S(2,1)$ arise in the bifurcation diagram
of Fig.~1(a) as functions of the parameter $\epsilon_2$. The
values of the instantaneous mean field of the wavelength in each
pattern $S(k,p)$, given by  $\langle f
\rangle_n=\frac{1}{k}\sum_{m=1}^k f(\sigma_n^m)$, are also plotted
on the corresponding regions of the bifurcation diagrams. In the
case of Fig.~1(a), we get $\langle f \rangle = [f(a)+f(b)]/2$,
independent of time, on the region $S(2,1)$. Note that at
$\epsilon_2=0.56$, the mean field $\langle f \rangle$ is equal to
the value of the drive $L_1=0.39$ in Fig.1(a). Since $\langle f
\rangle$ also corresponds to the mean field of the entire lattice,
it may be inferred that for the values of the parameters $r=2$,
$\epsilon_1=0.54$, and $\epsilon_2=0.56$, an autonomous coupled
system described by Eq.~(\ref{stcm}) with mean field global
coupling may also exhibit a spatiotemporal frozen pattern with
$k=2$ and $p=1$, while sustaining a constant value of its global
coupling function at $H=0.39$. For those parameter values, the
driven lattice Eq.~(\ref{drnw}) and the autonomous system
Eq.~(\ref{stcm}) are equivalent. Note that this prediction is
being made without direct numerical simulation on the autonomous
spatiotemporal system. Moreover, the intersection of $\langle f
\rangle$ with the constant value of $L_1$ in the diagram of the
driven lattice in Fig.~1(a) readily provides a set of initial
conditions for the odd and even elements $x_t^i$ in
Eq.~(\ref{stcm}) for observing the spatiotemporal pattern $S(2,1)$
in the autonomous coupled system at the parameter values
$\epsilon_1=054$, $\epsilon_2=0.56$, and $r=2$; that is,
$x_t^i=a=0.017$, for $i$ even; and $x_t^i=b=0.776$, for $i$ odd.
Similar predictions of parameter values and initial conditions for
the emergence of the spatiotemporal patterns $S(2,2)$, $S(2,4)$,
and $S(3,3)$ of examples (2)-(4) in the autonomous system
Eq.(\ref{stcm}) can be made from the intersections of the curves
$\langle f \rangle_n$ with the values of $L_t$ in the diagrams of
the periodically driven lattices of Fig. 1(b), 1(c) and 1(d),
respectively. Figs. 2(a)-2(d) display the spatiotemporal patterns
$S(k,p)$ of examples (1)-(4) emerging in the autonomous system
Eq.(\ref{stcm}) while sustaining either a constant or a periodic
mean field coupling function, at parameter values and orbits
predicted from the analogy with the driven coupled map lattice.

These results suggest that a complete
equivalence between a lattice driven with period $T$ (Eq.\ref{drnw}),
that gives rise to a spatiotemporal pattern $S(k,p)$  ($p/T=\nu$),
and an autonomous system (Eq.\ref{stcm}) can be established when
the following conditions are satisfied
\begin{equation}
\label{LH} L_n=
H(\ldots, \underbrace{\sigma_n^1,\sigma_n^2,\ldots,\sigma_n^k}_{N/k  \,
\mbox{times}} ,\ldots)
\end{equation}
where $n=1,\ldots,T$; and where the arguments of $H$ are the $k$ elements
in the $n$th row of
the matrix $S(k,p)$, repeated $N/k$ times.
For convenience, we denote
the right hand side of Eqs.~(\ref{LH}) by $H(\sigma_n^m)$.
For given parameters $r,\epsilon_1$,and $\epsilon_2$,
the orbits $\sigma_n^m$ of the driven lattice depend on
the periodic drive
$L_t=\{L_1,L_2,\ldots,L_T\}$, according to Eq.~(\ref{kl}). Thus,
Eqs. (\ref{LH}) constitute a set of $T$ nonlinear equations for the
values $\{L_1,L_2,\ldots,L_T\}$.
The solutions $L_t^*=\{L_1^*,L_2^*,\ldots,L_T^*\}$
of Eqs. (\ref{LH}) predict that the autonomous system (Eq.\ref{stcm})
with local and global interactions possesses a state characterized by
the spatiotemporal pattern $S(k,p)$ with orbits $x_n^m=\sigma_n^m(L_t^*)$, and
by the periodic global coupling function motion
$H=L_1^*, H=L_2^*, \ldots, H=L_T^*$.
As an illustration, consider the pattern
$S(2,1)$ whose orbits $a(r,\epsilon_1,\epsilon_2,L_1)$
and $b(r,\epsilon_1,\epsilon_2,L_1)$ can be obtained from Eqs.~(\ref{21}).
Then Eqs. (\ref{LH}) reduce to just one equation, which in the case of
a mean field coupling function $H=\sum_i^N f(x_t^i)$, yields the solution
\begin{equation}
L_1^*=\frac{1}{2r(\epsilon_2+2\epsilon_1-1)}.
\end{equation}
Thus for the set of parameters satisfying the above relation, the
pattern $S(2,1)$ can emerge in both, a steadily driven lattice
with $L_t=\{L_1^*\}$ and an autonomous system sustaining a
constant mean field coupling $H=L_1^*$.

Fig.~3 displays the  function $H(\sigma_n^m)$ corresponding to
mean field coupling
as a function of
the constant drive $L_1$ for the patterns of examples (1), (2) and (4).
The intersections of the curves with the diagonal give the solutions
$L_1^*$ to Eq. (\ref{LH}) for each pattern indicated.
Note that Eqs. (\ref{LH}) can be used to predict
if a given pattern $S(k,p)$ observed in  the driven
lattice may emerge in diverse autonomous systems
possessing different functional forms of the global coupling function $H$.
For instance, one may ask if the pattern $S(2,2)$ of example (2)
can also exist in an autonomous system Eq.~(\ref{stcm})
with a constant value of a global coupling function given
by the geometric mean, $H=\prod_i^N \left|x_t^i \right|^{1/N}$. Fig.~3
shows $H(\sigma_n^m)$ for this form of global coupling
associated to the pattern $S(2,2)$ (slash-dotted curve)
as a function of $L_1$, giving a solution
$L_1^*=H=0.3418$ at the intersection with the diagonal.
This prediction has been verified for this autonomous system.

In summary, we have studied the emergence of induced
ordered patterns in forced
spatiotemporal systems by using a model of
a one-dimensional coupled map lattice subjected to an external drive.
Under some circumstances, this system is analogous
to an autonomous coupled map system possessing
a similar local coupling and an additional global interaction
that acts as a global feedback.
Thus, by exploring the dynamical responses of
a driven spatiotemporal system, one can get insight into
the conditions for the emergence of specific patterns in a class of
autonomous spatiotemporal systems.
Once an ordered pattern $S(k,p)$ appears in a driven lattice for some
values of the parameters, the same pattern is expected to spontaneously arise
in a family of related autonomous coupled map systems that satisfy
Eqs.~(\ref{LH}), which constitute the link between both systems.
Conversely, a pattern observed in an autonomous system
with both local and global interactions should also exist in an associated
driven system with similar local couplings for some appropriate period of
the external drive.
This equivalence could be tested in experimental settings of interacting
dynamical elements \cite{Ring,Hudson2,Yamada,Hudson3}.
The method has been applied to some simple cases; however
it can be used for more complicated patterns.
Although we have considered one-dimensional diffusive local
couplings, the analogy between a uniform external drive
and a global interaction can be applied to any
network of coupled maps. The relation
between extended systems subjected to non-uniform driving fields
and spatiotemporal autonomous systems
is an interesting problem for future
research.

This work was supported by Consejo de Desarrollo Cient\'{\i}fico,
Human\'{\i}stico y Tecnol\'ogico of Universidad de Los Andes,
M\'erida, Venezuela.

\pagebreak

\begin{figure}
\epsfig{file=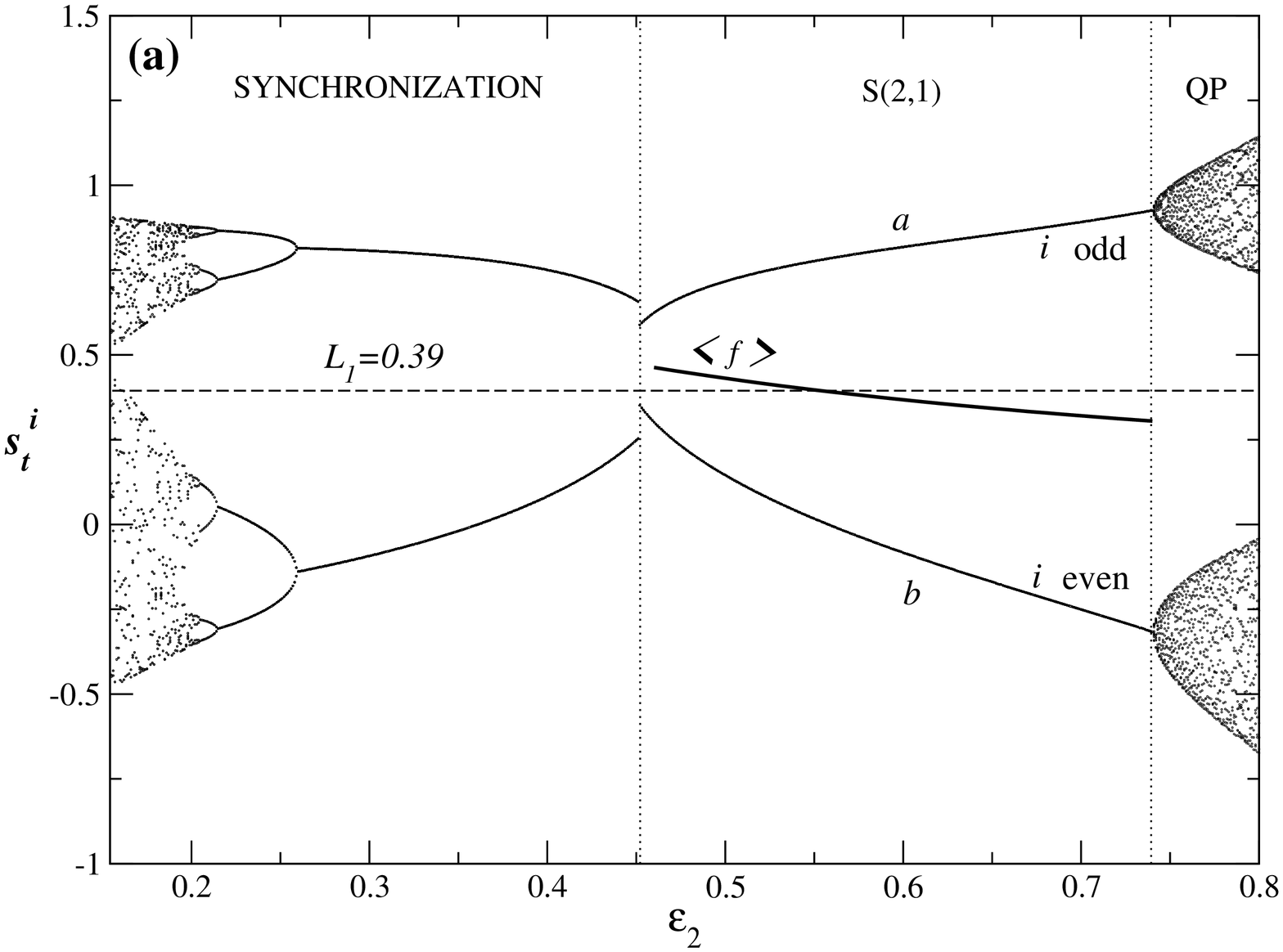,width=0.35\textwidth,angle=270,clip=}
\epsfig{file=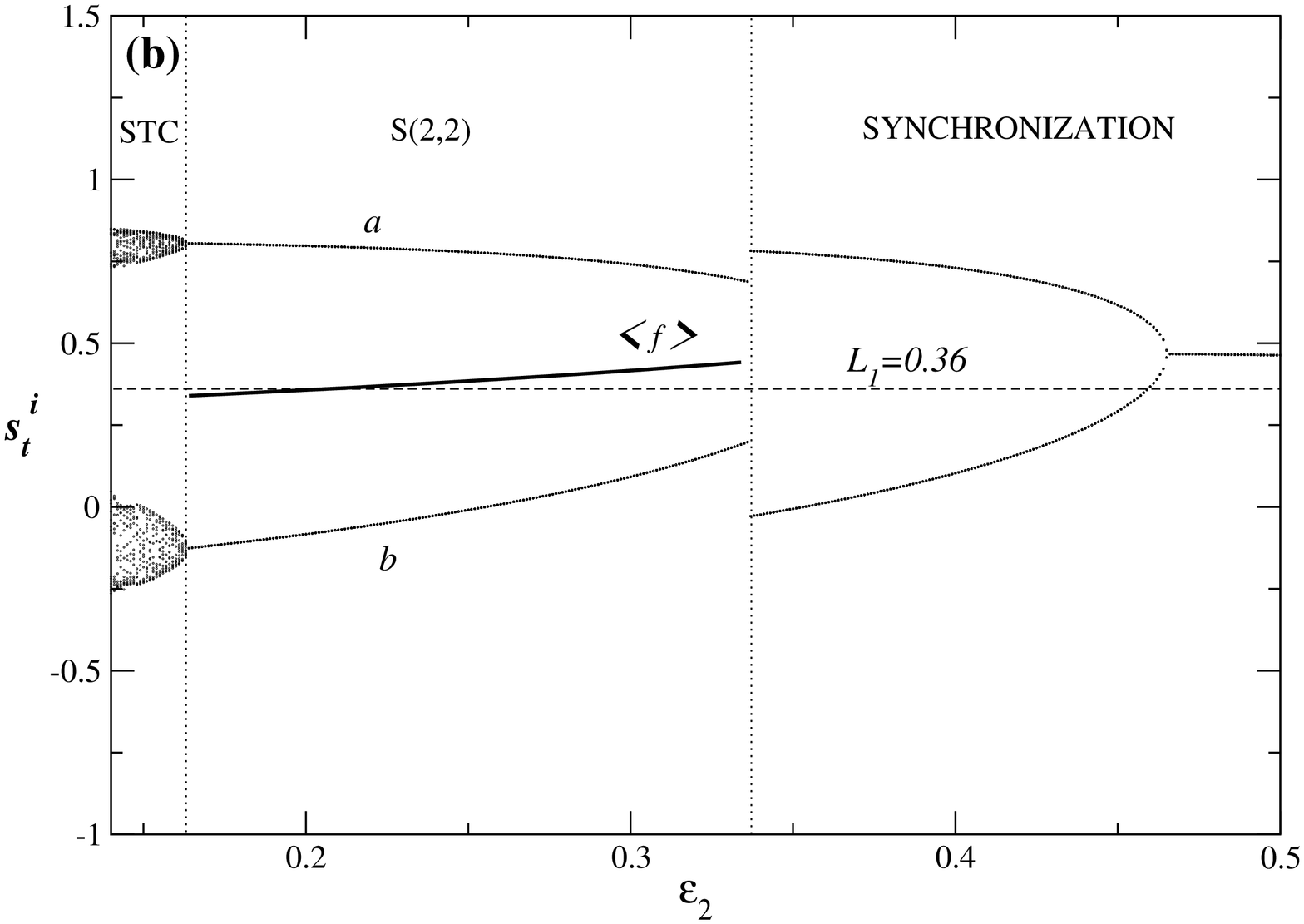,width=0.35\textwidth,angle=270,clip=}
\\
\epsfig{file=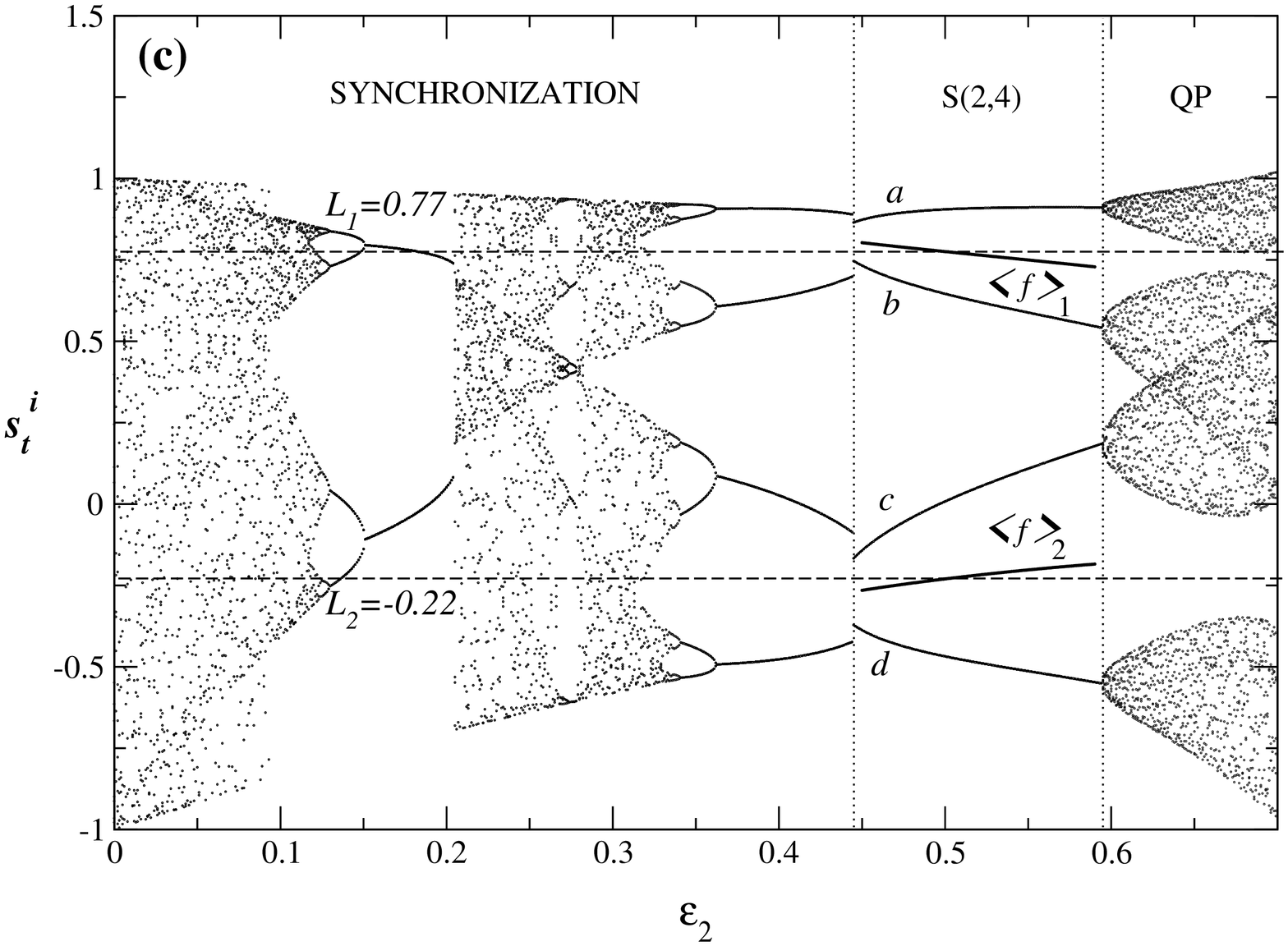,width=0.35\textwidth,angle=270,clip=}
\epsfig{file=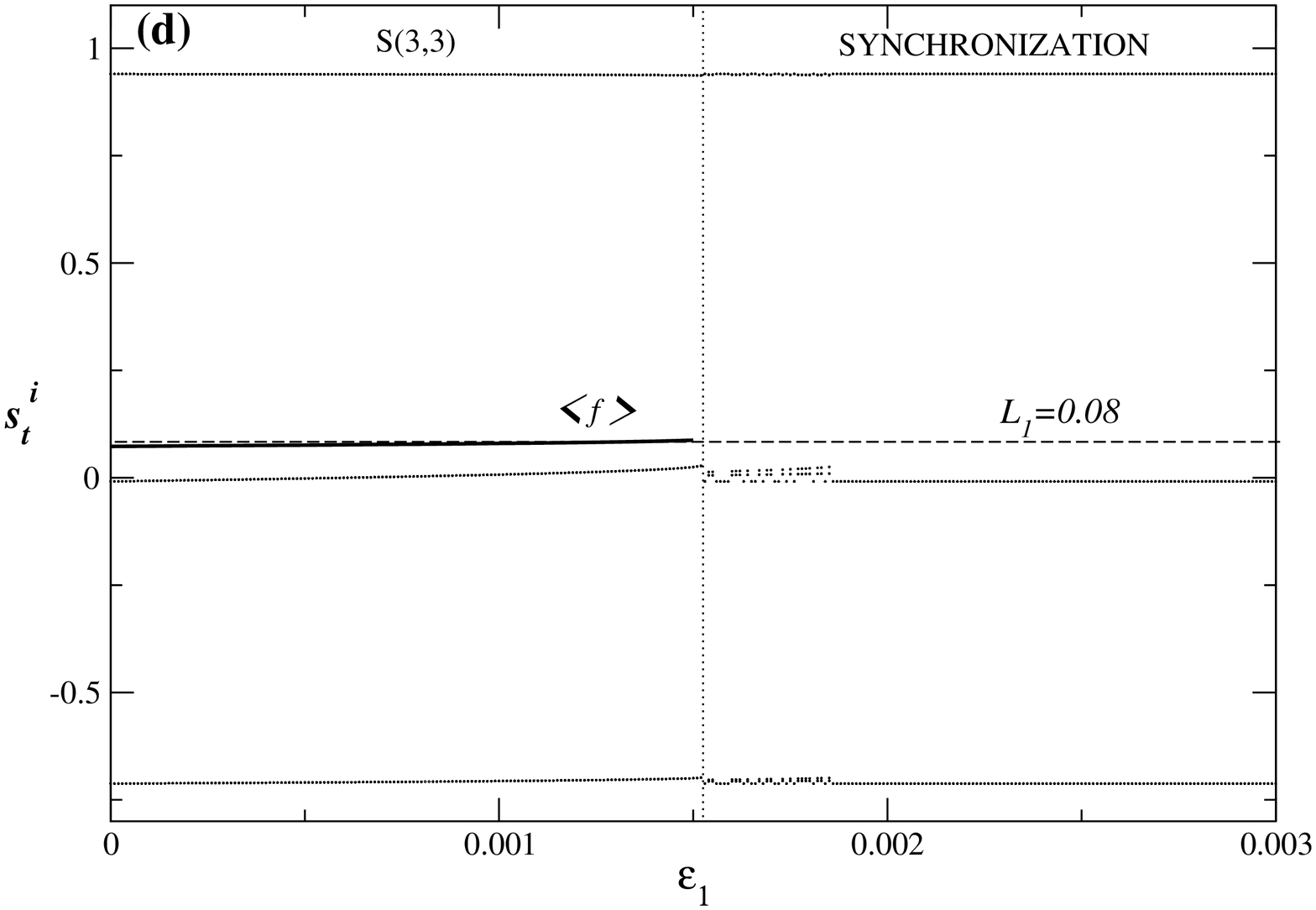,width=0.35\textwidth,angle=270,clip=}
\label{fig1}
\caption{Bifurcation diagrams of the orbits $s_t^i$ of the driven lattice
(Eq.\ref{drnw}) with size $N=30$
and fixed local parameter $r=2$. The values of the periodic drive $L_t$
are shown with dashed lines.  Orbits $a, b, c, d$,
corresponding to patterns $S(k,p)$ of examples (1)-(4) are indicated.
The mean $\langle f \rangle_n$ is drawn with
thick lines on each
region where a pattern $S(k,p)$ appears.
Regions where synchronization occurs
are identified; regions of quasiperiodic behavior are
labelled QP, and those of spatiotemporal chaos are labelled STC.
(a) $\epsilon_1=0.54$, and constant drive $L_1=0.39$; bifurcation parameter
is $\epsilon_2$.
(b) $\epsilon_1=0.05$,  $L_1=0.36$;  bifurcation
parameter $\epsilon_2$.
(c) $\epsilon_1=0.54$, and period-two drive $L_1=0.77, L_2=0.22$;
bifurcation parameter $\epsilon_2$.
(d) $\epsilon_2=0.065$;  $L_1=0.08$; bifurcation parameter is
$\epsilon_1$.
}
\end{figure}
\begin{figure}
\epsfig{file=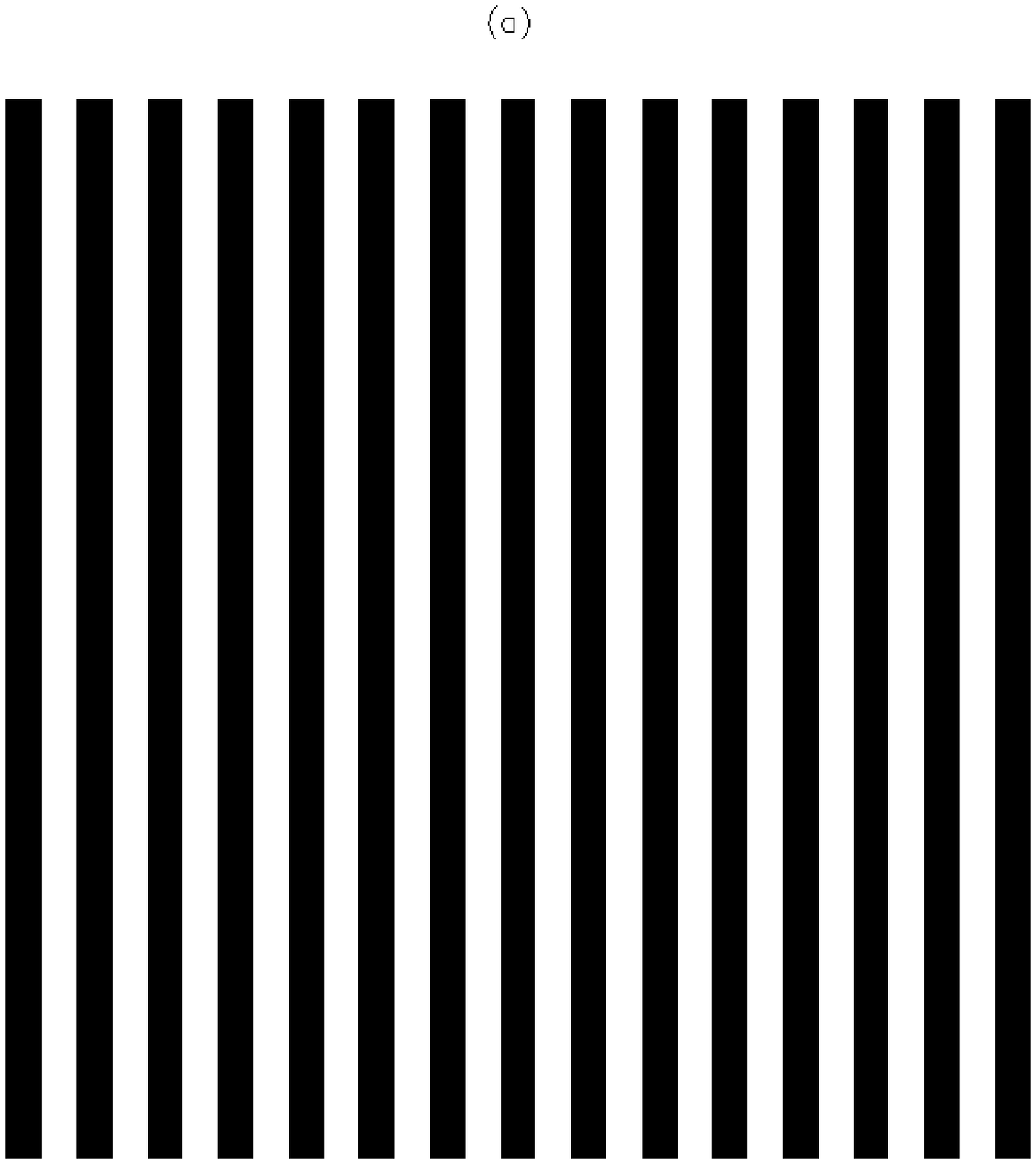,width=0.4\textwidth,angle=270,clip=}
\qquad
\epsfig{file=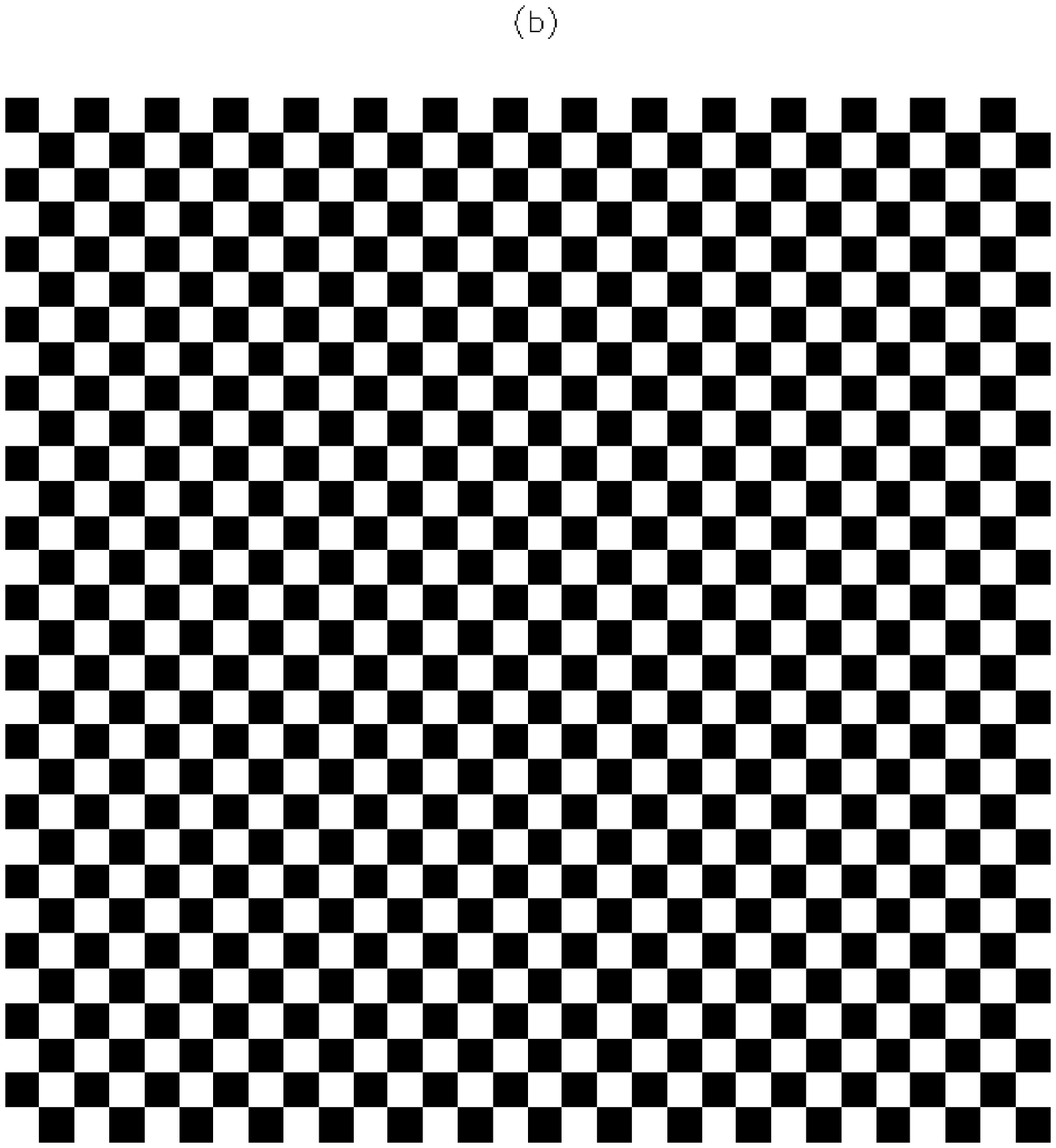,width=0.4\textwidth,angle=270,clip=}
\\
\epsfig{file=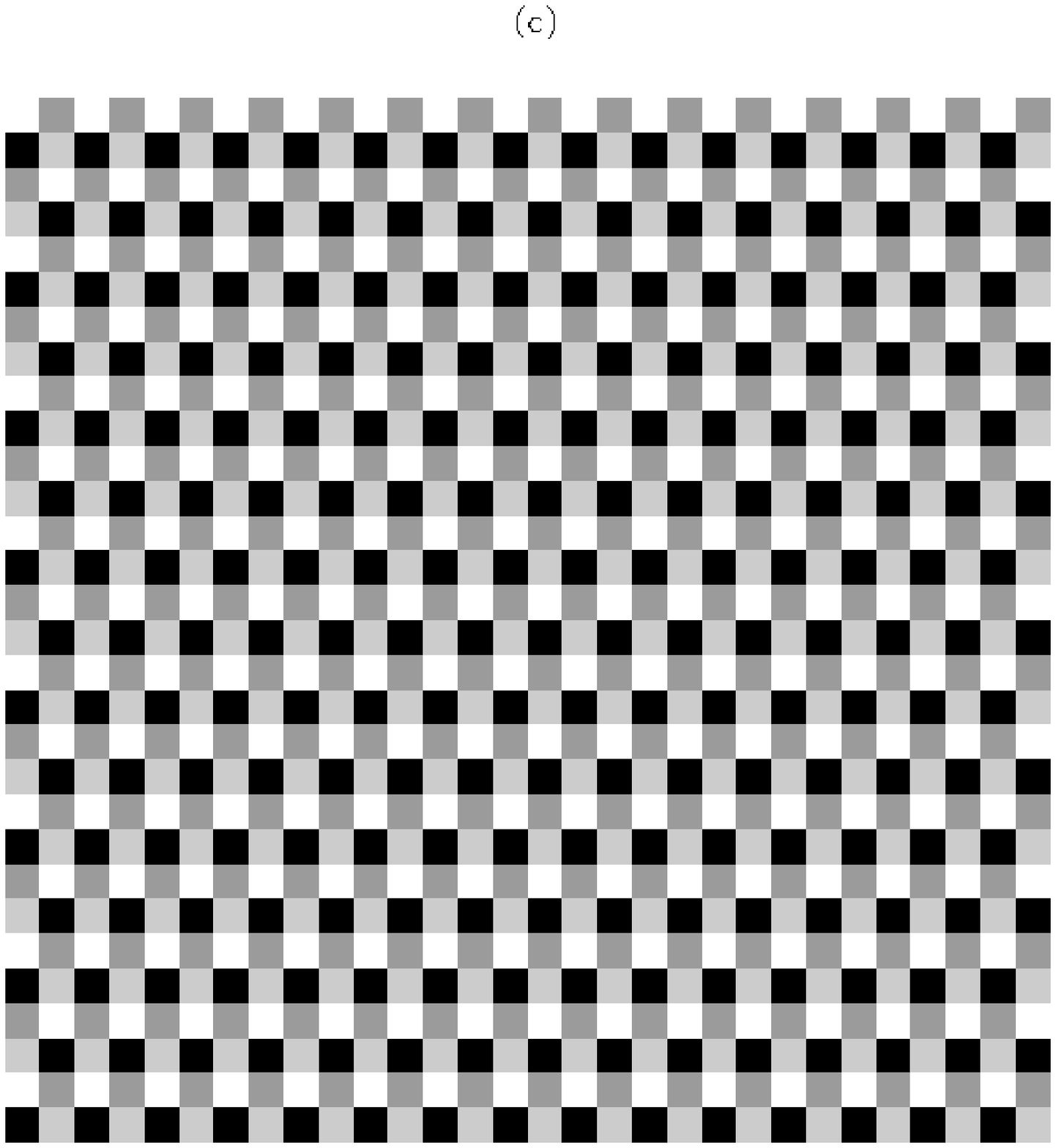,width=0.4\textwidth,angle=270,clip=}
\qquad
\epsfig{file=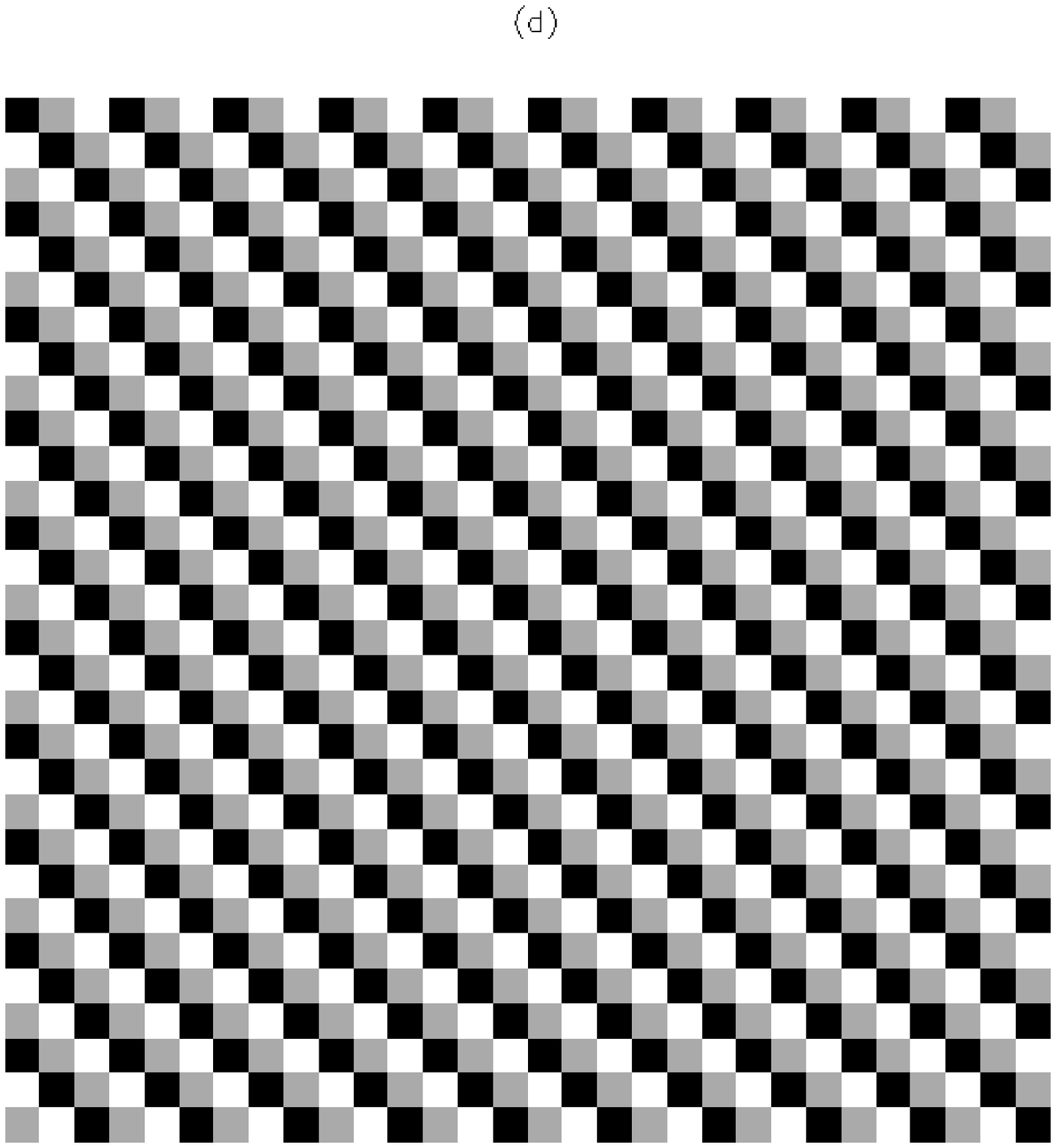,width=0.4\textwidth,angle=270,clip=}
\label{fig2}
\caption{Spatiotemporal patterns on a grey scale in the autonomous
system (Eq.\ref{stcm}) with
global mean field interaction $H$, at parameter values predicted from
the analogy with the driven lattice in Fig.~(\ref{fig1}). 
Size $N=30$; spatial
index of $x_t^i$ runs horizontally and time runs from bottom to top.
(a) Pattern $S(2,1)$; $r=2$, $\epsilon_1=0.54$, $\epsilon_2=0.56$;
constant $H=0.39$.
(b) Pattern $S(2,2)$; $r=2$, $\epsilon_1=0.05$, $\epsilon_2=0.2$;
constant $H=0.36$.
(c) Pattern $S(2,4)$; $r=2$, $\epsilon_1=0.54$, $\epsilon_2=0.51$;
$H$ oscillates periodically
between the values $0.77$ and $0.22$.
(d) Pattern $S(3,3)$; $r=2$, $\epsilon_1=0.001$, $\epsilon_2=0.065$;
constant $H=0.08$.
}
\end{figure}
\begin{figure}
\epsfig{file=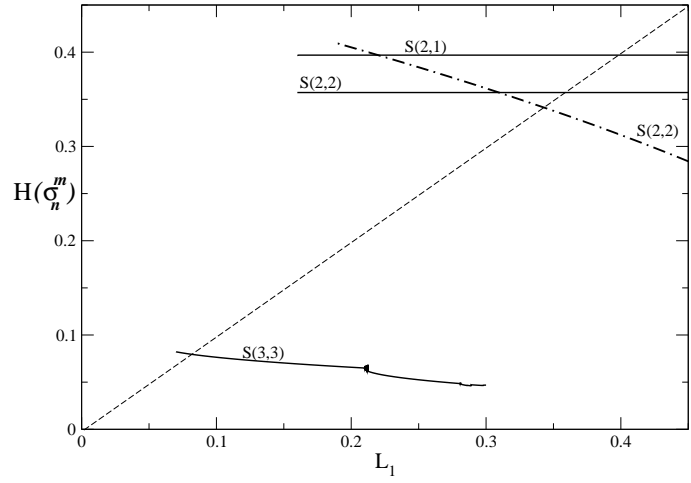,width=0.45\textwidth,angle=270,clip=}
\label{fig3}
\caption{Global coupling functions $H(\sigma_n^m)$ (r.h.s of Eq.~(\ref{LH}))
associated to patterns $S(k,p)$ with constant drive, as a function of
$L_1$. Continuous curves correspond to mean field global coupling
with the parameters of Fig.~(2) in each case.
Dash-dotted curve displays a geometric mean global coupling function
associated to the pattern $S(2,2)$ with parameters
$r=2, \epsilon_1=0.005, \epsilon_2=0.4$; the intersection
occurs at $L_1^*=0.3418$.
}
\end{figure}
\end{document}